\documentclass[submission,copyright,creativecommons]{eptcs}
\providecommand{\event}{GramSec 2014} 

\usepackage[T1]{fontenc}%
\usepackage{url}
\usepackage{amssymb}
\usepackage{graphicx}
\usepackage{algorithm}
\usepackage{algorithmic}
\usepackage{color}
\usepackage{listings}
\usepackage{multirow}
\usepackage{hyperref}
\hypersetup{
    bookmarks=true,         
    unicode=false,          
    pdftoolbar=true,        
    pdfmenubar=true,        
    pdffitwindow=false,     
    pdfstartview={FitH},    
    pdftitle={My title},    
    pdfauthor={Author},     
    pdfsubject={Subject},   
    pdfcreator={Creator},   
    pdfproducer={Producer}, 
    pdfkeywords={keyword1} {key2} {key3}, 
    pdfnewwindow=true,      
    colorlinks=false,       
    linkcolor=red,          
    citecolor=green,        
    filecolor=magenta,      
    urlcolor=cyan           
}

\usepackage{color}
\title{Towards the Model-Driven Engineering of Secure yet Safe Embedded Systems}

\author{Ludovic Apvrille
\institute{Institut Mines-Telecom, Telecom ParisTech, CNRS LTCI\\ Sophia Antipolis, France}\\
\email{ludovic.apvrille@telecom-paristech.fr}
\and
Yves Roudier
\institute{EURECOM\\Sophia Antipolis, France}\\
\email{\quad yves.roudier@eurecom.fr}
}

\def\titlerunning{SysML-Sec}
\def\authorrunning{L. Apvrille, Y. Roudier}

\begin{document}
\maketitle

\begin{abstract}
We introduce SysML-Sec, a SysML-based Model-Driven Engineering environment aimed at fostering the collaboration between system designers and security experts at all methodological stages of the development of an embedded system. A central issue in the design of an embedded system is the definition of the hardware/software partitioning of the architecture of the system, which should take place as early as possible. SysML-Sec aims to extend the relevance of this analysis through the integration of security requirements and threats. In particular, we propose an agile methodology whose aim is to assess early on the impact of the security requirements and of the security mechanisms designed to satisfy them over the safety of the system. Security concerns are captured in a component-centric manner through existing SysML diagrams with only minimal extensions. After the requirements captured are derived into security and cryptographic mechanisms, security properties can be formally verified over this design. To perform the latter, model transformation techniques are implemented in the SysML-Sec toolchain in order to derive a ProVerif specification from the SysML models. An automotive  firmware flashing procedure serves as a guiding example throughout our presentation.

\end{abstract}

\section{Introduction}
Most contributions around Model Driven Engineering (MDE) now offer appropriate methodologies and modeling environments for designing safe, complex, distributed, and real-time embedded systems. The analysis of timing constraints, scheduling, resource allocation, and concurrency are commonly handled by these environments. In contrast, security has long been considered in retrospect, especially after serious flaws were discovered in computerized systems. Security as well as privacy issues have in particular only recently become a major concern in embedded systems. However, the size, heterogeneity, and communication features of modern embedded systems make it compelling to develop a suitable engineering environment to more explicitly define security objectives and threats, to implement countermeasures with security mechanisms, and to assess or even formally prove the effectiveness of security countermeasures.

We introduce SysML-Sec, a SysML-based Model-Driven Engineering environment aimed at fostering the collaboration between system designers and security experts at all methodological stages of the development of an embedded system. SysML-Sec introduces both customized SysML diagrams for security matters and an associated methodology. The SysML-Sec methodology includes three SysML-based stages. (i) System analysis starts with a  partitioning-based process in which  security requirements and threats can be identified together with functional features of the system. (ii) System design focuses on software-implemented security mechanisms. Finally, (iii) System validation intends to formally verify, simulate, and test the models built at previous stages by relying on model transformation techniques. This paper presents the overall methodology, with a particular focus on the design and proof of security mechanisms.

The SysML-Sec methodology and diagrams have been developed and experimented in the scope of the FP7 European project EVITA, which resulted in the design and implementation of a secure architecture for automotive embedded systems. The definition, design, and validation of this architecture was performed with the methodology that is presented in this paper. Thus, more than 20 use cases (notably an emergency braking use case) were taken into account for that purpose. The diagrams in this paper are directly excerpted from the EVITA "firmware flashing" case study.

\label{sec:introduction}

\section{Context: Embedded Systems}
\label{sec:context}
\subsection{Designing Embedded Systems}
IT systems are commonly designed following a V-cycle, with building stages (requirements, analysis, design, deployment) followed with verification stages (e.g., tests, formal proofs). For embedded systems, the V-cycle can obviously start only once functions have been partitioned into software and hardware.
System partitioning usually relies on the Y-chart approach \cite{FB-POLIS-03}. This is the very first step to co-design software and hardware functions on the one hand, and the hardware architecture (defined in terms of execution, communications, and storage) on the other hand.
The result of this process is an optimal hardware / software architecture with regards to criteria at stake for that particular system (e.g., cost, performance,  etc.). In the scope of the DIPLODOCUS environment \cite{LA-ICECS06}, on top of which SysML-Sec is built, the Y-Chart is implemented as follows:
\begin{enumerate}
\item {\it Applications}  are first described as abstract communicating tasks: tasks represent functions independently from their implementation form. 
\item Hardware {\it architectures} are described as a set of abstract execution nodes (e.g., CPU with operating systems and middleware, hardware accelerators), communication nodes (e.g., buses), and storage nodes (e.g., memories). 
\item A {\it mapping} model \cite{FB-POLIS-03} defines how tasks and communications between tasks are assigned to computation and communication / storage elements, respectively and partitioned between hardware and software. For example, a task mapped on a hardware accelerator is a hardware-implemented function whereas a task mapped over a CPU is a software implemented function.
\end{enumerate}

This partitioning process is of utmost importance. Indeed, if critical high-level design choices are  invalidated afterwards because of late discovery of issues (performance, power, etc.), then it may induce prohibitive re-engineering costs and late market availability.\\
DIPLODOCUS relies on the SysML allocation mechanisms for the mapping stage. The UML deployment diagram might be a good candidate for this, but it is rather used for the deployment of already-designed software functions: "the assignment of software artifacts to nodes", as stated in the UML standard. In DIPLODOCUS, functions may be fully hardware implemented. Moreover, they are highly abstracted, that is, we do not map any concrete artifacts (e.g., a source file), but only high-level functional elements.

\subsection{Security Issues in Embedded Systems}

An increasing number of embedded systems have become communicating artifacts, feature new interactions with their immediate environment or with backend systems, and are thus exposed to criminals. For example, attacks have been shown to be possible on set-top boxes like Microsoft's XBox \cite{Huang02} or ADSL routers \cite{Assolini12}, mobile appliances \cite{Esser11}, avionics  \cite{Teso13}, or automotive systems
\cite{Hoppe08}
to cite but a few.
 Many of these security issues reflect either the exploitation of low-level vulnerabilities, which might often be addressed with appropriate programming practices and specific component tests, or design flaws due to an insufficient understanding of the mapping of functional or security logical components to the hardware architecture. We claim that the SysML-Sec Model-Driven Engineering approach makes it possible to perform an appropriate system analysis, design and proof in both directions, and to describe both security threats and security objectives and to further prove whether their are well handled at system design.

\section{SysML-Sec: an Overview}
\label{sec:methodology}
\subsection{Rationale}
We designed the SysML-Sec environment in order to make it possible to describe security issues together with partitioning requirements, as further discussed in \cite{MoDRE13}. In particular, our extensions bridge the gap between goal-oriented descriptions of security requirements and attacks, and the fine-grained representation of assets based on the software / hardware architecture (and their model-driven analysis). SysML-Sec also supports phases of the V-cycle after the partitioning stage, and notably the design of the software-partitioned functions\footnote{SysML-Sec does not address hardware design}. The main objectives of SysML-Sec are:
\begin{itemize}
\item Guiding and increasing the collaboration between system engineers and security experts throughout the entire embedded system lifecycle. This has been the reason for our adoption of the OMG standards, and more specifically SysML, which are quite widespread in the embedded system world today.
\item Providing detailed representations of the security threats and security requirements compatible with the MDE methodology used and making it possible to adopt a stepwise refinement approach to the definition of both the functional and the security architecture. This refinement should also make it possible to bridge the gap between initial high-level requirements and the definition of precise and detailed security mechanisms later on.
\item  Combining software/hardware codesign together with the handling of security concerns. We contend that this particular design objective is a key in the embedded system domain.
\item Offering simulation and formal verification capabilities at system partitioning, system design, which constitute two critical phases of embedded system engineering. At the system partitioning level, simulations help assess the impact of security features on the system performance, e.g., the impact of a security protocol on a bus load. At the system design level, formal verification intends to prove whether threats are correctly handled.
\end{itemize} 

\subsection{Methodology}

\begin{figure}[htbp]
\centering
\includegraphics[scale=0.5]{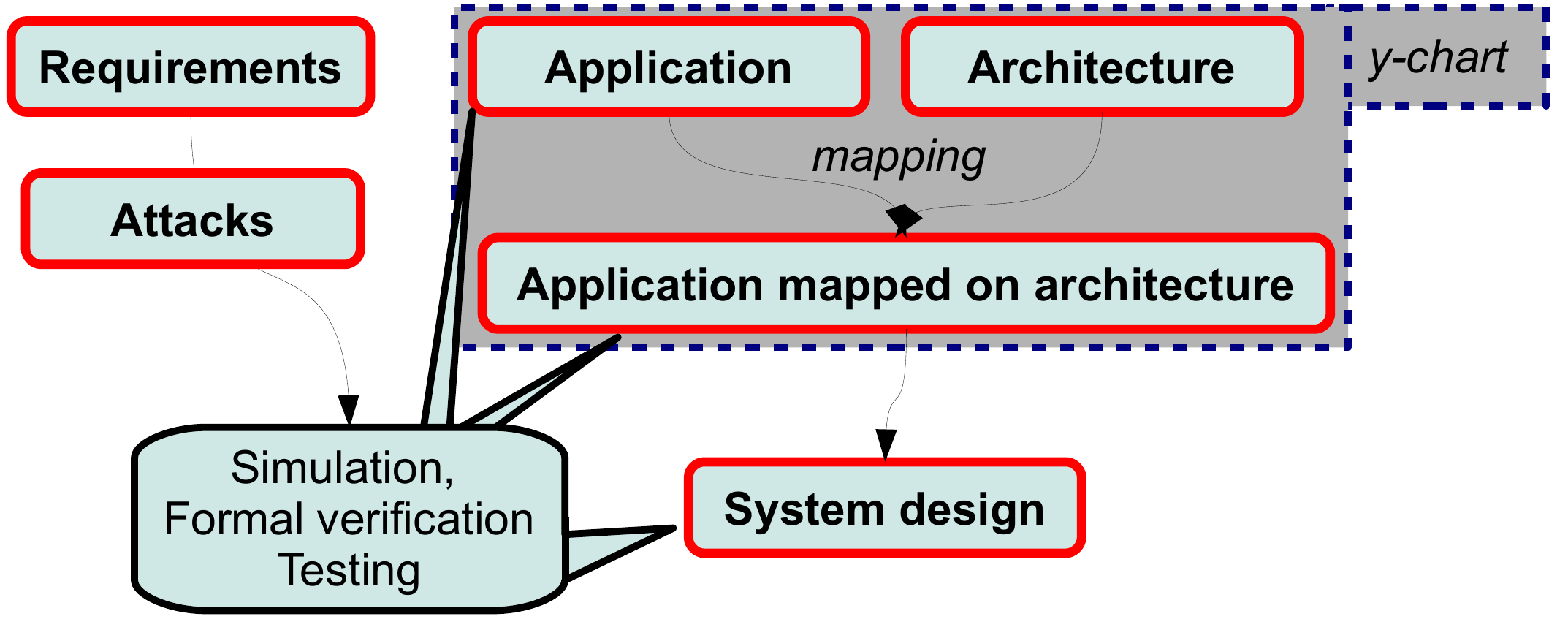}
\caption{SysML-Sec methodology} \label{fig:sysmlsec_methodo}
\end{figure}

The SysML-Sec methodology adopts a three-phase approach that first deals with the system analysis, then with software design, and finally with system validation, as depicted in Figure~\ref{fig:sysmlsec_methodo}. 
The analysis includes the elicitation of requirements and attacks, and the partitioning of the system. The design stage includes the definition of security mechanisms, and the refinement of security requirements in security properties to be proved in the design. Lastly, the verification takes place at different engineering phases, as follows. Simulation is mostly used at the partitioning stage in order to evaluate the impact of security mechanisms in terms or performance.  Formal verification intends to prove the resilience of the system under design to threats. Testing is meant to do the same, but on the deployed implementation resulting from the design models.

\subsection{Tooling}
TTool \cite{ttool} is a free software supporting several UML profiles, e.g., DIPLODOCUS and AVATAR. For the partitioning stage, SysML-Sec reuses DIPLODOCUS, as explained in \cite{EURECOM+3484}. Requirements, attacks, and design are analyzed with AVATAR, a profile dedicated to the analysis and modeling of embedded systems \cite{LA-WI13}.

\section{Case Study: Firmware Update}
\label{sec:casestudy}
We illustrate SysML-Sec methodological stages with a "firmware update" case study. The latter is taken from a public deliverable of the European project EVITA \cite{EvitaD2.1}. The main purpose of EVITA is to define a secure architecture for automotive embedded systems. Such systems are in charge of critical functions and are quite complex: they contain around 100 Electronic Controls Units (ECUs) all interconnected to the main system bus (CAN, FlexRay). Attacks are motivated by safety and economical reasons (notably theft). Security concerns, which were mostly related with physical access to the car are also evolving today with the advent of the connected car and of Car2X communications, which further exposes these systems.

The considered case study aims at the update of an ECU software with a newer version of the firmware, as performed at a service station through diagnosis tools. Car diagnosis is hardwired according to the Standard Unified Diagnosis Services UDS, which is specified in the ISO 14229-1.
At first, the ECU to be flashed initializes its software and starts the diagnosis function. The service station employee connects his diagnosis tool to the on-board diagnosis interface in the vehicle. Once the diagnosis has been performed and authentication is performed, a programming session is settled. Finally, the flashing can be performed in ROM.

The overall process is expected to be secure, in particular with respect to integrity, confidentiality, and authenticity of the firmware and/or of the flashing process. More specifically, the following security requirements can be defined (see \cite{EvitaD2.1} for further details):
\begin{itemize}
\item \textbf{Authenticity}: is the vehicle sure to communicate with a valid diagnosis Tool?
\item \textbf{Confidentiality}: the firmware constitutes intellectual property to protect.
\item \textbf{Data Integrity}: the flashed code must not have been modified.
\item  \textbf{Anonymity}: Private information about the driver should not be disclosed to the service station during the flashing process.
\end{itemize}

\section{System Requirement Engineering and Analysis}
\label{sec:analysis}
The security requirement and threat analysis is mostly regarded as a preamble to risk analysis in IT systems. This process is generally meant to decide whether to introduce security countermeasures into the system, which means additional costs. In the case of embedded systems, we contend that the security analysis also has a strong impact on the  system architecture and its realtime performance: the security requirements and threat analysis should thus be performed along an iterative, and therefore more agile partitioning process.

\subsection{Iterative Security/System Codesign Process}

System partitioning, security requirements, and threats are progressively refined based on one or several typical use cases. The following phases, which thus start with an initial architecture, are iterated in order to reach a satisfactory level of refinement:

{\bf Initial Architecture Mapping.} The functionalities of the system highlighted in these use cases are first modeled as tasks. Exchanges between functions are modeled with information and event flows between tasks. Tasks and communications can  then be mapped to a draft architecture of the system. The designer's experience plays a key role in determining the first draft of the architecture.

{\bf Architecture Analysis}
{\it System assets} are identified among architectural elements (processors, pieces of software, sensors, hardware accelerators, communication channels) and will first refer to generic components, like for example: "all system buses". When the architecture gets more detailed, assets are more likely to be refined into specific elements. The hardware/software partitioning and the function mapping adopted play a key role here in defining the type of asset at hand (and later on its vulnerabilities).

{\bf Security Concern Identification.} 
{\it Threats and security vulnerabilities of the selected assets} should as much as possible describe the capabilities that an attacker should meet or exceed and the origin of attacks (local, remote, through a specific interface). The SysML-Sec environment supports the assessment of risks following the approach described in more detail in the EVITA case study \cite{EvitaD2.3,EVITA-req-risk}. We also implemented automated checks of the threat coverage by security objectives. Based on the risk analysis, one should also identify and prioritize security objectives that are mapped to a threat.

{\it Security Objectives} might originate (1) from security standards or properties expected from the system, or (2) from unaddressed threats or attacks on assets, or (3) from the refinement of another security objective when the process is iterated and the level of detail of the architecture has changed. In further iterations, one may need to update security objectives deprecated by changes in the architecture.

{\bf Architecture Refinement.} The architecture refinement originates from a more detailed description of the architecture components as the system and its usage become more precisely known (e.g., new communication channels, refinement of an execution environment into OS/middleware/application layers, etc.). It may also result from transitively mapping requirements to system information flows, which are often distributed among multiple hardware elements. The refinement phase may fail if the architecture and security requirements are incompatible, for instance, if the performance overhead of security mechanisms is too high. Consistency checks should also be performed to ensure that a security objective does not conflict with another requirement expressed over the same asset. A failure is the sign that the analysis should be backtracked to the previous stage of refinement.

\subsection{Diagrams}

\subsubsection{Requirements}

Security requirements are modeled in SysML Requirement Diagrams (RD). The main operators of  RDs are \textit{Requirement Containment} and \textit{Derive Dependency} formalisms used to define relationships between requirements.  The \textit{containment} relationship depicts sub-requirements in terms of hierarchy and enables a complex requirement to be decomposed into its containing child requirements whereas \textit{deriveReqt} determines the multiple derived requirements that support a source requirement. 
A \textit{Security Requirement} stereotype is introduced to make a clear distinction between functional requirements and security requirements of the system, yet modeling both functional and non-functional requirements in a single environment. Furthermore, a \textit{Kind} parameter is defined to specify the category of the security requirement (\textit{confidentiality}, \textit{access control}, \textit{integrity}, \textit{freshness}, etc.).

\begin{figure*}[htbp]
\centering
\includegraphics[scale=0.7]{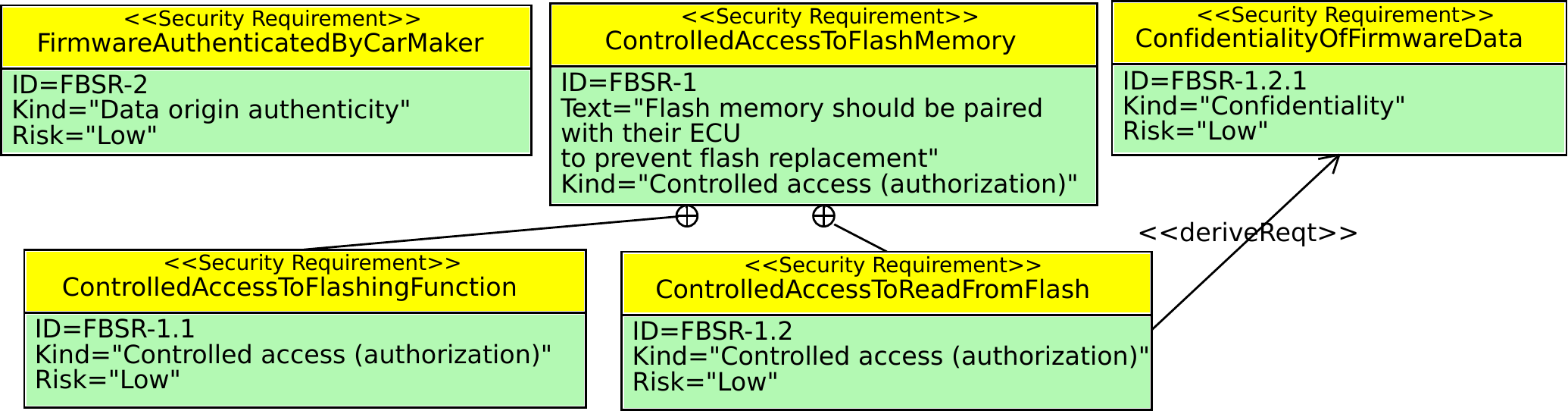}
\caption{Excerpt from the SysML-Sec Requirement Diagram of "Firmware Update"} \label{fig:furd}
\end{figure*}

Figure \ref{fig:furd} represents 5 security requirements for a "Firmware update": the authenticity of the firmware, controlled access to the flash memory, itself derived into controlled access to both the flashing function and to reading the flash, and the confidentiality of firmware data.

\subsubsection{Threats and Attacks}
Instead of using the traditional attack tree approach \cite{schneier}, we suggest that threats can be better modeled with a more relational approach, using slightly customized SysML Parametric Diagrams. Threats are modeled as values embedded into blocks representing the target of the attacks, thus achieving a representation that is more compact and better mapped to the system architecture. Attacks ($<<attack>>$ stereotype) can be linked together with logical operators like $OR$, $AND$, as well as temporal causality operators like $SEQUENCE$, $BEFORE$, or $AFTER$. We consider the latter constructs as especially helpful to describe the attacker's operational point of view in embedded systems, like for instance situations in which there is a maximum duration between two causally related attacks. For example, when attacking a system with time-limited authentication tokens, the token must be first retrieved, and then the use of this token must occur before its expiration.

Attack instances in different parametric diagrams can be linked together in order to assess the impact of a specific vulnerability and the need to address it at the risk assessment phase. 
An attack can also be tagged as a $root$ attack, meaning that this attack is at the top of a tree of attacks. Last but not least, attacks can be linked to requirements, thus allowing an automated check of the coverage of attacks.

\begin{figure*}[htbp]
\centering
\includegraphics[scale=0.7]{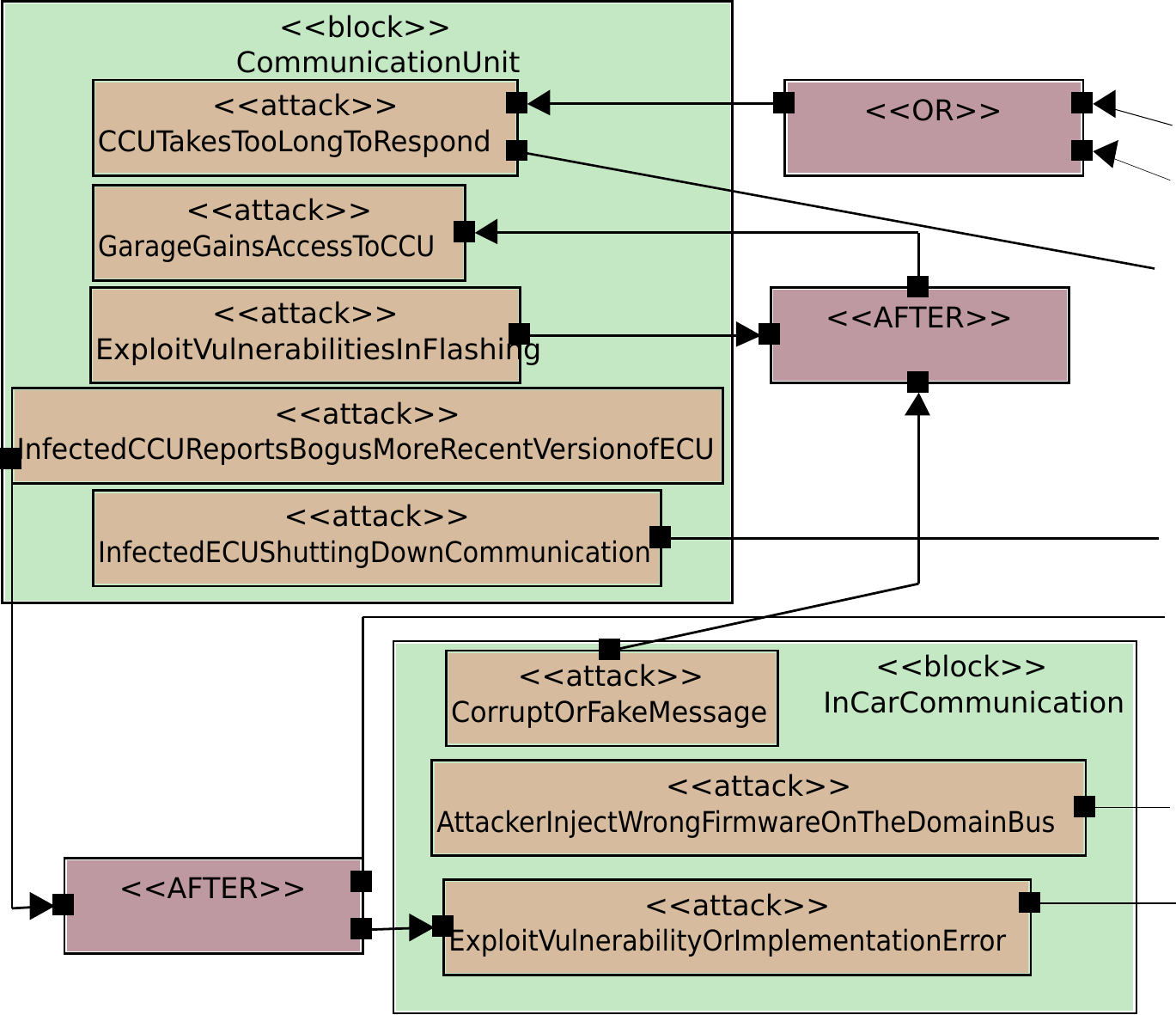}
\caption{Excerpt from the Attack Diagram of "Firmware Update"} \label{fig:fuat}
\end{figure*}

Figure \ref{fig:fuat} depicts a few attacks identified in the scope of the EVITA "Firmware update" case study. Two assets are represented (\textit{CommunicationUnit}, \textit{InCarCommunication}). For example, performing the attack "ExploitVulnerabilityinFlashing" in \textit{CommunicationUnit} and then forging a  "CorruptOrFakeMessage" in the \textit{InCarCommunication}  makes it possible to perform "GarageGainsAccessToCCU".

\begin{figure*}[htbp]
\centering
\includegraphics[scale=0.58]{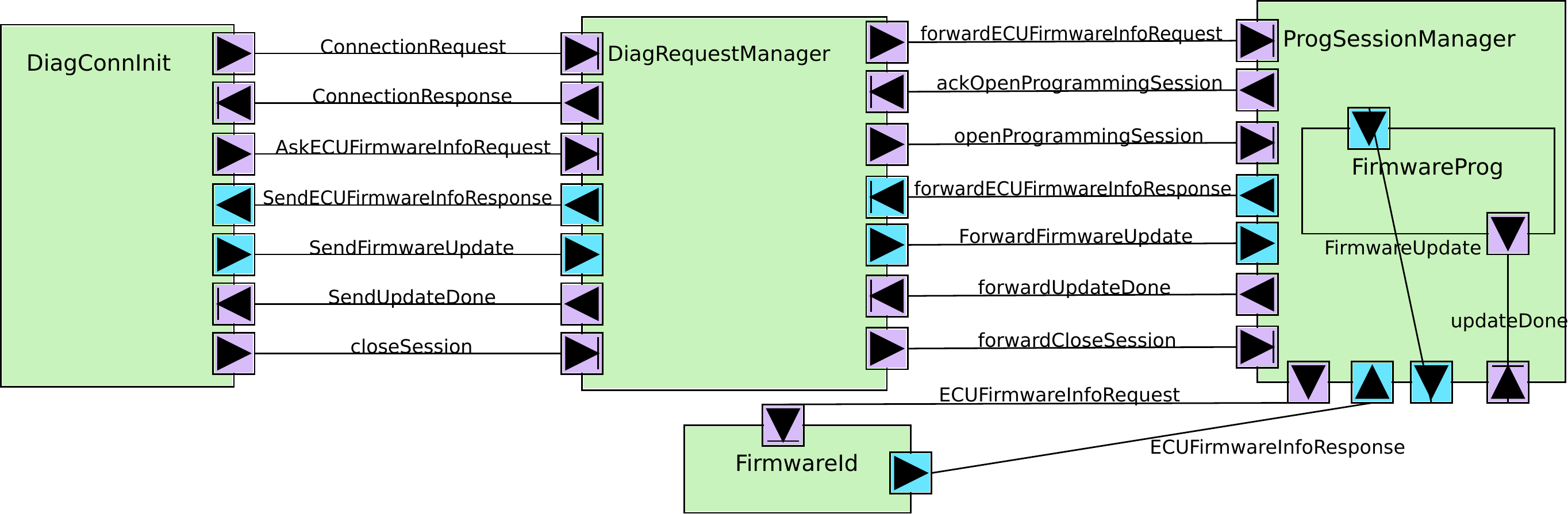}
\caption{Application model of firmware update} \label{fig:fv_fu}
\end{figure*}

\subsubsection{Partitioning}

The application model is a graph of communicating functions. For example, in Figure \ref{fig:fv_fu}, event flows correspond to purple ports, and information-flows to blue ports. Five main functions are modeled: the initialization of the diagnosis connection (\textit{DiagConnInit}), diagnosis request management (\textit{DiagRequestManager}), programming session management (\textit{ProgSessionManager}), firmware identification (\textit{FirmwareId}) and lastly, the firmware programming (\textit{FirmwareProg}).

\begin{figure*}[htbp]
\centering
\includegraphics[scale=0.7]{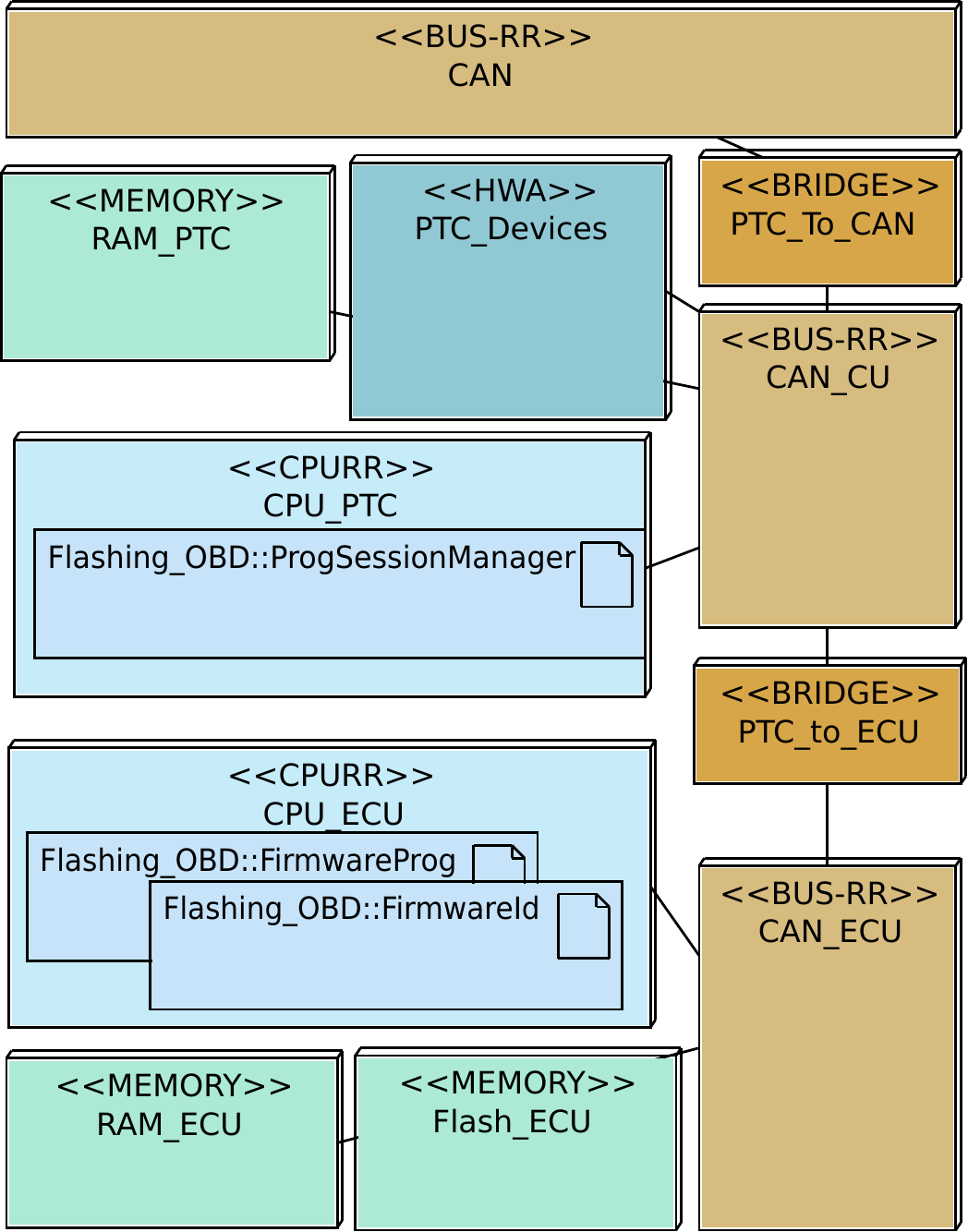}
\caption{Excerpt from the architecture and mapping model of firmware update} \label{fig:mapping_fu}
\end{figure*}

The architecture model is partially depicted in Figure \ref{fig:mapping_fu}. It represents the assets to be protected. Two subdomains are connected to the main CAN bus. Each subdomain has its own CPU, RAM, and bus. The first domain also has a hardware accelerator, and the second one has a flash memory that the procedure intends to update. 

The mapping consists in assigning functional elements to assets, i.e., assigning tasks to either a CPU or hardware accelerators, and communications to buses and memories. Figure \ref{fig:mapping_fu} displays the mapping of three tasks in the two subdomains: \textit{ProgSessionManager} is mapped to "CPU\_PTC" and \textit{FirmwareProg} and \textit{FirmwareId} are mapped on "CPU\_ECU".

\section{System Design}
\label{sec:design}
\subsection{Methodological Aspects}
System design intends to refine the architecture and behavior of all functions mapped over processor nodes during the partitioning. From a security point of view, the aims is to describe in more detail how security requirements can be fulfilled with security hardware mechanisms of software mechanisms executed on top of the hardware architecture defined in the partitioning stage, and to verify that requirements identified during the partitioning phase are really satisfied by this design. Requirements expressed at partitioning are informal and refer to assets: they therefore need to be refined until their expression directly relates to design elements (e.g., attributes, methods, exchanged messages, states, etc.).  Once refined, they constitute the security properties that are to be verified, notably through formal methods.

Diagrams of SysML generally used for system design, like block diagrams and state machine diagram, lack explicit ways to model security mechanisms. For example, security mechanisms commonly need the pre-sharing of cryptographic material (e.g., secret keys), but block diagrams have no way to model this. Thus,  SysML-Sec extends SysML in order to explicitly model security mechanisms and properties.

\subsection{Security Design Extensions}
\begin{figure*}[htbp]
\centering

\begin{tabular}{cc}

\includegraphics[scale=0.65]{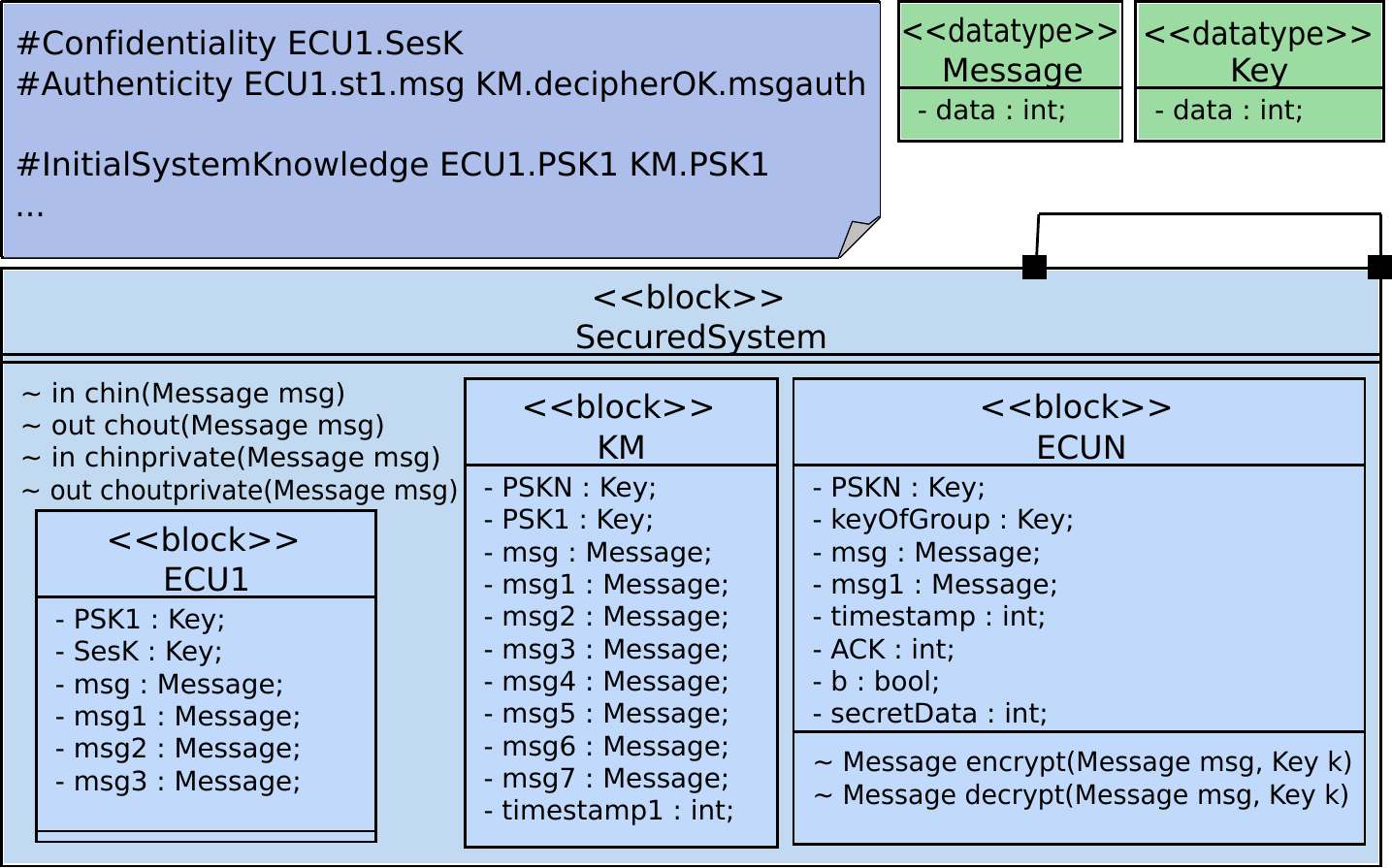}
&
\includegraphics[scale=0.58]{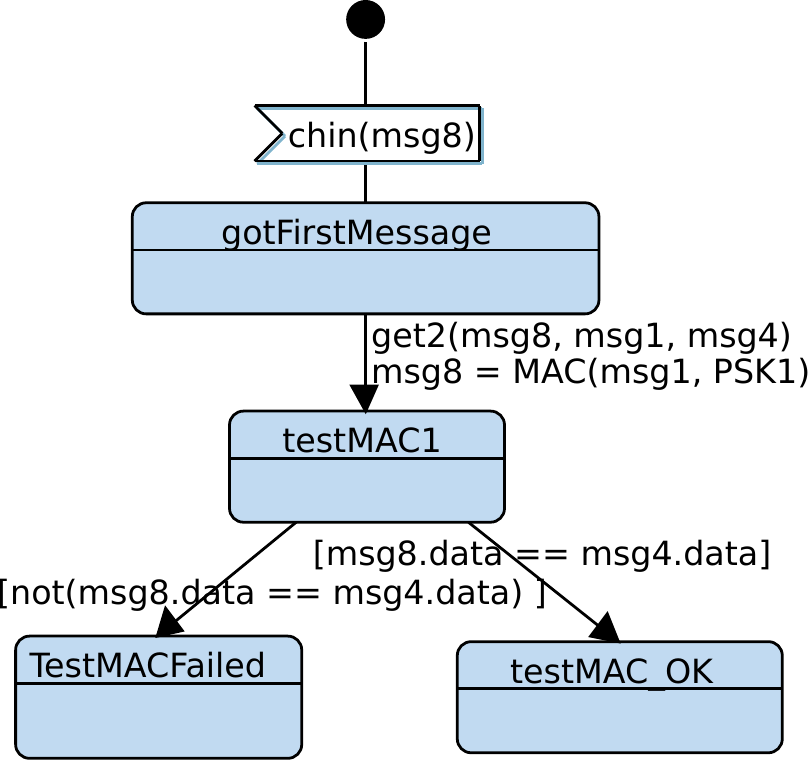}
\end{tabular}

\caption{SysML-Sec block diagram and a state machine diagram of Key Distribution Protocol} \label{fig:smd_kmp} \label{fig:bd_kmp}
\end{figure*}

A SysML-Sec design is made upon SysML block and state machine diagrams, extended with several features, and formally defined in pi-calculus (a process algebra).
We assume a Dolev-Yao attacker model, that is, only  messages exchanged between blocks can be eavesdropped, contrary to attributes of blocks that are considered as private from the point of view of the attacker. That attacker model is enough to describe attacks on the protocols deployed between the components of the embedded system, from outside (i.e., using communication networks) or from within the system (i.e., using internal buses or any other accessible component interface within the system). It however does not aim at capturing physical attacks on the hardware, nor a sequence of exploitation of vulnerabilities of several components.
The main extensions are:
\begin{itemize}
\item \textbf{Public and Private Channels}: Since communication channels may have been mapped over secure or non secure buses at the partitioning stage, we give the possibility to tag links between blocks with a $public$ label if an attacker can eavesdrop, or with a $private$ label otherwise.
\item \textbf{Cryptographic Algorithms}. SysML-Sec blocks can define a set of methods corresponding to cryptographic algorithms (e.g., \textit{encrypt}), \textit{verifyMAC}, etc. so as to be able to describe security mechanisms built upon these algorithms, e.g., cryptographic protocols.
\item \textbf{Cryptographic Material}. Blocks can also pre-share values, a feature commonly needed to setup cryptographic protocols. SysML-Sec introduces specific \textit{pragmas} for that purpose: ({\it InitialSystemKnowledge} and {\it InitialSessionKnowledge}). The first one is used to describe data that are shared before the system execution. The second one defines data that are shared within each session of the same system. Typically, when considering a cryptographic protocol, the first pragma means that the data are pre-shared and common to all protocol sessions, and the second one states that the data are shared but have different values in each  protocol session.
{\small
\begin{flushleft}
\texttt{$\sharp$ InitialSystemKnowledge BlockID.attribute} $[$\texttt{BlockID.attribute}$]$*
\end{flushleft}
}
\end{itemize}
For example, Figure \ref{fig:bd_kmp} displays the block diagram of a key distribution protocol used as a security protocol so as to allow ECUs to communicate in a secure way. That protocol is built upon three different entities: the initiator of the key distribution (\textit{ECU1}), a key master (\textit{KM}) and the ECUs with which ECU1 expects to share keys (\textit{ECUN}). All blocks define cryptographic functions, two of them being visible in ECUN (\textit{encrypt()}, \textit{decrypt()}). The pragma "InitialSystemKnowledge" states that \textit{PSK1} is an attribute which is shared between ECU1 and KM before system startup. 

Figure \ref{fig:smd_kmp} presents the subset of the KM state machine diagram. State machines are used to describe the behavior of each block. They can manipulate cryptographic functions and data. For example in the state machine of the KM block, the action $msg8=MAC(msg1, PSK1)$ assigns the result of $MAC(msg1,PSK1)$ to $msg8$ .

\subsection{Security Properties}
Security properties linked to the design are obtained from a refinement of security requirements elicited at analysis stage. A dedicated language has been defined for describing the commonly complex safety properties, which is based on SysML Parametric diagrams \cite{DK-SIGSOFT-11}. On the contrary, security properties can usually be defined with a type (e.g., \textit{confidentiality}), and with design elements related to that kind (e.g., the confidentiality of the attribute of a block). This simplicity pleads for a basic modeling solution, that is not based on complex diagrams or operators. Our solution again relies on \textit{pragmas}, provided in notes of Block Diagrams: \textit{confidentiality} and \textit{authenticity} can be directly expressed at this level.

A confidentiality pragma states that the content of an attribute of a block shall never be disclosed to an attacker:\\
{\small
\texttt{$\sharp$ Confidentiality block.attribute}\\
}

An authenticity pragma states that a message $m2$ received by a block $block2$ was necessarily sent before by block $block1$ in a message $m1$.
The following examples describes such a situation:\\
{\small
\texttt{$\sharp$ Authenticity block1.s1.m1 block2.s2.m2}\\
}

 This authenticity pragma specifies two states: one of the sender block, i.e. one state $s1$ in $block1$, and one state $s2$ in $block2$. Also, in the state machine diagram of $block1$, $s1$ corresponds to the state right before the sending of  $m1$. Analogously, $s2$ corresponds to the state  right after message $m2$ has been received and accepted as authentic.

The confidentiality pragma of Figure \ref{fig:bd_kmp} states that the PSK1 (pre-shared key 1) of ECU1 shall never be disclosed (secret key). This property is derived from the requirement on confidentiality: secret keys must remain secret so as to ensure the confidentiality of the firmware that is to be sent to the flash memory. Similarly, the authenticity pragma states that \textit{msgauth} received by KM right after state \textit{decipherOK} has necessarily been sent by ECU in variable \textit{msg} right before state \textit{st1}.

\section{System Validation}\label{sec:validation}
A system validation can be performed from partitioning models (e.g., performance evaluation of the selected hardware architecture: load of CPUs and buses), from design models (e.g., proof of safety and security properties), or from executable code automatically generated from deployment models (e.g., safety and security tests) \cite{LA:ERTSS-12}.
Model transformations have been defined in order to transform SysML-Sec models into simulation, formal, or executable specifications.  Mapping model transformations are given in \cite{phd_knorreck}, and design model transformations are described in \cite{phd_pedroza} and \cite{these-Sabir}.

From partitioning models, it is possible to evaluate the impact of security mechanisms onto safety constraints, e.g., real-time constraints such as latencies. An example of such a study is given in \cite{HS-WIV-11} where the impact of adding security-oriented MAC information in the messages of CAN bus is checked against safety-oriented properties (latencies) at partitioning stage. The impact of security mechanisms onto the load of buses and CPUs might be evaluated as well, as they might impact the behaviour of safety-critical processes and communications. For instance, the decryption of packets sent from an ECU to another one may prevent the scheduling of the process controlling the brake actuation.

\begin{figure*}[htbp]
\centering

\begin{tabular}{cc}

\includegraphics[scale=0.8]{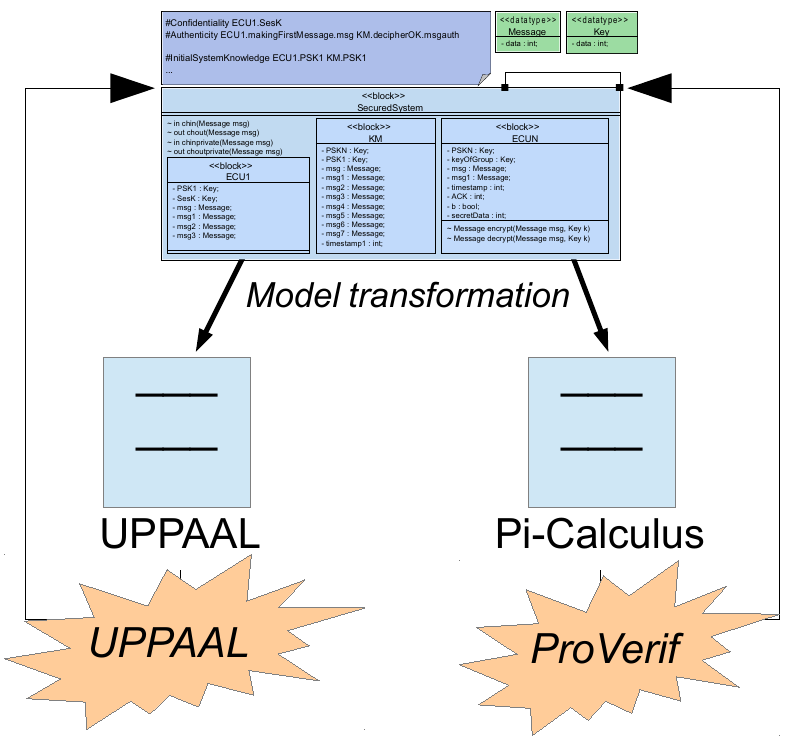}
&
\includegraphics[scale=0.5]{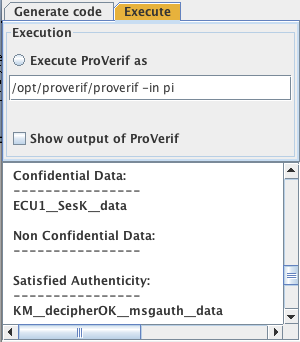}

\end{tabular}

\caption{(a) Model transformations of TTool for proving safety and security properties. (b) TTool assistant for the formal verification of confidentiality and authenticity properties defined at Figure \ref{fig:bd_kmp}}

\label{fig:proof}
\label{fig:proverif}

\end{figure*}

From SysML-Sec designs, the formal proof relies
respectively on UPPAAL \cite{uppaal-paper} and  
on \textit{ProVerif} \cite{BlanchetJCS08} for the proof of safety and 
security properties (see Figure \ref{fig:proof}). Our safety proofs take into account all design elements, including security mechanisms like the liveness and reachability of all states of cryptographic protocols, and the impact of security mechanisms onto safety-critical processes. 
Our security proofs may relate to any behavior described with state machines. The notion of protocol is central here: it captures both communications over buses or networks and communications between components in separate blocks like separate trust zones within a processing unit, for instance when virtualization mechanisms are used. 

\textit{ProVerif} \cite{BlanchetJCS08} is a toolkit that relies on Horn clauses resolution for the automated analysis of security properties over cryptographic protocols, under the Dolev-Yao model. ProVerif takes in input a set of Horn Clauses, or a specification in pi-calculus together with a set of queries. ProVerif outputs whether each query is satisfied or not. In the latter case, ProVerif tries to identify a trace explaining how it came to the conclusion that a query is not satisfied. Figure \ref{fig:proverif} depicts the successful verification in TTool of the confidentiality and authenticity properties modeled in Figure \ref{fig:bd_kmp}. While we can specify any security requirement, we currently only support the formal validation of those that can be expressed with these two security properties. 

Safety proofs take into account all design elements apart from the ones specifically defined for security purposes, that is the security-oriented pragmas and cryptographic methods that have no impact on safety properties (liveness, reachability). Similarly, the proof of security properties abstracts away irrelevant system details. Such a proof does not require values on variables nor temporal information. Several state machine elements, like temporal operators (\textit{after} clause) and computations with variable values, are not taken into account for the security proof. For instance, $a=b+c$ is executed in a symbolic way, but without concrete values: the data flow captured in that manner (we know that $a$ contains information from $b$ and $c$) is the only useful information for security proofs.

\section{Related Work and Perspectives}
\label{sec:relatedWork}

Many proposals already exist that address security requirements engineering or threat modeling in IT systems.
In \cite{Nhlabatsi2010}, Nhlabatsi et al. classify security requirements engineering work according to four dimensions, namely: {\it goal-based approaches}, {\it Model-based approaches}, {\it problem-oriented approaches}, and {\it Process-oriented approaches}.
Our approach combines a goal-based description of security requirements with a model-driven engineering of the system architecture and threat analysis, and emphasizes the importance of agile interactions throughout the engineering V-cycle.
In that respect, its philosophy follows that of the TwinPeaks approach \cite{Nuseibeh}, even though the latter does not address hardware systems. Instead of a simple spiral alternating between the requirements and the architecture as TwinPeaks suggests, we also alternate between the Y-Chart modelling of software and its mapping to hardware components, the identification of assets and threats to them, and the identification of security requirements. 

Assessing security properties at design level mostly relies on formal approaches. For example,  \cite{MJToussaint} proposes to verify cryptographic protocols with a probabilistic analysis approach.
In more recent efforts,  \cite{QSIC2004} introduces a first order Linear Temporal 
Logic (LTL) into the Isabelle/HOL theorem prover, thus making it possible to model 
both a system and its security properties, but unfortunately leading to non-easily reusable specific models. \cite{Amana.GPujol} mixes formal and informal security properties, but the overall verification process is not completely automated, again requiring specific skills.
Our approach focuses on refining semi-formal specifications into formal models using a graphical language, with the aim of formal validation of the satisfaction of security properties.
In this respect, it may seem very similar to the UMLsec approach \cite{dses07},
a modeling framework aimed at defining security properties of software. UMLsec also primarily features a model-based approach to security requirements engineering according to the abovementioned classification. It makes it possible to define security properties of software components and of their composition within a UML framework. It also features a rather complete framework addressing various stage of model-driven secure software engineering from the specification of security requirements to tests, including logic-based formal verifications regarding the composition of software components.
With respect to the embedded system field which the current paper focuses on, UMLsec allows for the description of the mapping of already-designed software onto hardware nodes \cite{dses07} through the use of UML deployment diagrams. According to OMG's own definition, such diagrams describe ``a set of constructs that can be used to define the
execution architecture of systems that represent the assignment of software artifacts to nodes.''. This means that they unfortunately cannot handle software-hardware partitioning in the sense of \cite{FB-POLIS-03}. 
In contrast, SysML-Sec adopts the more holistic point of view of SysML proponents and focuses on the very evaluation of various mappings for security and functional features before and during detailed system design. Additionally, SysML-Sec provides hardware nodes with an explicit autonomous simulation and proof semantics. Lastly, SysML-Sec includes low-level hardware nodes whose use may impact communications and executions of functions, e.,g. a DMA engine may favorize, or not,  a urgent communication with regards to security-oriented communications.
Similarly to UMLsec\cite{uusgtssd02}, our proposal additionally features a goal-based approach to describe security objectives. Still, our approach fits entirely within SysML with minor extensions (essentially new stereotypes and a few operators) and does not introduce any new diagram in order to make it easy to adopt for system engineers.

From our experience, partitioning is in our opinion a central issue in embedded systems. Achieving a correct partitioning that adheres to safety requirements necessitates that the impact of security mechanisms be understood and quantified as early as possible. We note that only a few authors, notably Eames and Moffet \cite{DBLP:conf/safecomp/EamesM99}, and more recently Pi\`etre-Cambac\'ed\`es \cite{DBLP:journals/ress/Pietre-CambacedesB13} and Raspotnig \cite{DBLP:journals/jss/RaspotnigO13}, have dealt with the relationships between security and safety requirements. For instance, the last two authors discuss quite thoroughly the relationships that can be established between security and safety requirements. In particular, these studies can describe conflicts, like we considered in the current paper, but also reinforcement relationships (when safety and security concur towards the same design), or conditional dependence. We think that obtaining similar descriptions within SysML-Sec would require the engineering methodology to be extended with an additional feedback interaction from all engineering phases to the specification phase: for instance, the satisfaction of safety requirements should be checked based on the security mechanisms introduced before any further safety mechanism would be introduced. We did not evaluate any such methodological step in our case study, which did not feature safety requirements. However, we plan to investigate such issues in the future.

The work we presented in this paper aims more specifically at validating the innocuity of security mechanisms with respect to safety requirements whose specification is outside the scope of this paper. This is based on quantitative validations during the specification and design phases, together with the refinement of the system architecture. In particular, approximations can first be made based on the nature of security mechanisms, and most notably, the cryptographic algorithms used and their costs (both computationally and in terms of bandwidth usage). As the design of the security mechanisms to be deployed is increasingly refined, together with a more detailed hardware/software partitioning, simulations by the SysML-Sec framework can be used to achieve a more precise evaluation of those elements. Although not yet supported by our toolchain, tests might finally be used on the developed components to validate the adherence to safety requirements.
Threat analysis itself pleads for such a move, to incorporate the description of security tests performed over implementations into the system model.
In this respect, our proposal includes the use of the SysML parametric diagram for threat modeling. SysML-Sec can thus encode an attack tree, yet it adopts a block-centric perspective with reuse in mind. We especially think that this will allow for the composition of the threat modeling performed by security analysts about components over-the-shelf (COTS) with system specific analyses. We plan to further extend SysML-Sec expressivity here: our declarative approach should be especially useful in order to incorporate knowledge from other threat modeling approaches.

Tool support is in our opinion critical to the successful and systematic engineering of systems, and especially so in what regards non-functional requirements like security and the validation of their satisfaction in a given design. This need has been recognized by many other authors at various levels of the development lifecycle of IT systems. Furthermore, the availability of free software is another important factor of success, would it only be due to their extensibility for specific needs. The KAOS framework \cite{VanLamsweerde2007}, which pioneered goal-oriented security requirements engineering, features an entire dedicated environment. CARiSMA (a recent extension of UMLsec) integrates into the Eclipse development environment, which ensures a better acceptance factor. Threat modeling frameworks have also been developed like Isograph Software's AttackTree+ for designing attack trees or the free software ADTool \cite{Kordy}. There also exists a number of formal verification tools with nice interfaces like AVISPA \cite{AVISPAtool}, which provides good support for protocol verification. It can be observed that many of these tools address only a part of the development lifecycle of a system, which is detrimental to their adoption. Our framework aims at a more comprehensive environment adhering to SysML, a formalism supported by the OMG and INCOSE, and widely known to system engineers.

\section{Conclusion and Future Work}
\label{sec:conclusion}

Embedded systems are becoming ever more complex and intricate combinations of software and hardware, exposed to a plethora of attacks as those systems open up to attackers through the introduction of communication functionalities. The economic incentives to reduce the time-to-market of such systems while ensuring equally good or better engineering qualities require the introduction of systematic security engineering methodologies with an appropriate support. The adoption of user-friendly tools featuring automated and simplified verification will likely become mandatory in this respect. 
Our proposal, SysML-Sec, specifically addresses that need at the diverse phases of system design and development.
It is based on a popular language (OMG's SysML) and supported by a free software toolkit (TTool). It features formal verification capabilities that rely on recognized toolkits for security (ProVerif) and safety (UPPAAL). 
The methodology we developed has been experimented to define a secure automotive embedded system. We presented the results of a case study on that system, which highlights several advantages of our approach, especially regarding formal verification.
Our future work will be dedicated to further evaluating our approach on other embedded systems as well as to extending the tool support, especially for enhancing threat coverage and requirements traceability analyses.

\bibliographystyle{eptcs}

\bibliography{gramsec14}
\label{sec:biblio}

\end{document}